# Quasi-optimal observables *vs.* event selection cuts

F.V.Tkachov

*Institute for Nuclear Research of Russian Academy of Sciences*
*Moscow, 117312, Russian Federation*

The method of quasi-optimal observables [hep-ph/0001019] offers a fundamental yet simple and flexible algorithmic framework for data processing in high energy physics to improve upon the practice of event selection cuts.

No better preface could be imagined for my presentation of the **method of quasi-optimal observables** [1] than the talk we've just heard [2]: there is obviously a number of HEP data processing specialists around the globe seeking ways — similarly to Dr. Anipko et al. — to extract maximum signal from their event samples. The talk [2] is also another evidence (not that there is a lack thereof) that the prescriptions which I am going to discuss are unknown to many physicists (see [3] for a discussion of this; see, however, *Notes added* hereafter).

I am going to advocate the method of quasi-optimal observables as a systematic solution of the problem of optimal HEP measurements, including but not limited to, the finding of optimal event selection cuts. The method:

- derives from first principles of mathematical statistics;
- is simple and fundamental (and as such deserves to be known to every non-mathematical physicist);
- offers a rather universal guiding principle for doing data processing in HEP, especially in the precision measurement type problems;
- offers a flexible algorithmic framework that seems to require neither artificial intelligence, nor neural networks (not to mention *ants*); in fact its implementations can be based on available adaptive algorithms for multidimensional integration (among other options) and seem otherwise straightforward.

The method came about as a by-product of the theory of jet definition [4] and is a result of attempts to understand the basics of HEP data processing under a premise that adding a fancy idea to a mess (a customary way of doing things in what is known as *normal science* [5]) usually creates only more mess (my favorite illustration of the law is the design of C++; see [6] and refs. therein).

Let $\{\mathbf{P}_i\}_i$ be the events (instances of a random variable $\mathbf{P}$) distributed according to the probability density $\pi(\mathbf{P})$; the density is assumed to depend on a parameter denoted as $M$. Take any textbook of mathematical statistics for physicists and recall that the simplest method of parameter estimation[1] is that of (generalized) moments going back to Pearson (1894) where one fixes a function on events (weight) $f(\mathbf{P})$ and then finds the experimental estimate for $M$ — denote it as $M[f]$ — by fitting the theoretical mean value, $\langle f \rangle = \int d\mathbf{P}\, \pi(\mathbf{P}) f(\mathbf{P})$, against its experimental analog $\langle f \rangle_{\text{exp}} = N^{-1} \sum_i f(\mathbf{P}_i)$; denote as $\mathbf{Var}\, M[f]$ the variance of the estimate, then the actual error from the fit is estimated as $\sqrt{N^{-1} \mathbf{Var}\, M[f]}$. The method is simple but is considered inferior to the extent that it is no longer mentioned in the PDG booklets. The problem is that there have been no prescription to choose $f(\mathbf{P})$ sensibly, i.e. so as to minimize $\mathbf{Var}\, M[f]$.

On the other hand, Fisher's (1912) method of maximal likelihood prescribes to estimate $M$ by the value which maximizes the expression $\sum_j \ln \pi(\mathbf{P}_j)$. The method may be difficult to apply (especially if the probability density is not known analytically or the number of events is large, as is often the case in HEP), but it is optimal in the sense that the resulting estimate $M_{\text{opt}}$ corresponds to the absolute lower bound as established by the fundamental Fisher-Frechet-Rao-Cramer inequality which for our purpose can be written as follows:

$$\mathbf{Var}\, M[f] \geq \mathbf{Var}\, M_{\text{opt}}. \qquad (1)$$

But however inferior the method of moments may seem, it represents a fundamental viewpoint in that the probability distribution can be *equated* to the collection of all average values $\langle f \rangle$ (recall the mathematical definition of distributions as linear functionals on test functions; probabilistic measures are special cases of distributions). However abstract this may seem, at least it's an indication (it was for me) that the method of moments may have a deeper significance than is the popular perception.

So, let us ask a simple question: given that $\mathbf{Var}\, M[f]$ is bounded from below, where in the space of $f$ is the minimum located? A baffling aspect of all this (discussed in [3]) is that there is little evidence that in the 100+ years since the introduction of the method there have been serious attempts to find ways to determine which $f$ are better than others in this respect — given that the minimum has been known to exist for 50+ years (see, however, *Notes added* hereafter).

Once one's asked the question, it is straightforward to write down the following criteria for the minimum:

$$\frac{\delta}{\delta f(\mathbf{P})} \mathbf{Var}\, M[f] = 0, \qquad \frac{\delta^2}{\delta f(\mathbf{P}) \delta f(\mathbf{Q})} \mathbf{Var}\, M[f] > 0. \qquad (2)$$

Further, in the statistical limit one has

$$\mathbf{Var}\, M[f] = (\mathbf{Var}\, f) \times [\partial \langle f \rangle / \partial M]^{-2}. \qquad (3)$$

Simple calculations (for details see [1]) yield the solution:

$$\boxed{f_{\text{opt}}(\mathbf{P}) = \frac{\partial \ln \pi(\mathbf{P})}{\partial M}} \qquad (4)$$

(up to an additive and a multiplicative constants that may both depend on $M$) and that for small deviations from optimality

$$\mathrm{Var}\, M[f_{\text{opt}} + \varphi] = \langle f_{\text{opt}}^2 \rangle^{-1} + O(\varphi^2) \langle f_{\text{opt}}^2 \rangle^{-3} + \ldots \qquad (5)$$

where $O(\varphi^2)$ is a known non-negative expression, which fact is essentially equivalent to the FFRC inequality (1). Note that $f_{\text{opt}}$ has to be fixed for some $M_0$. If $M_1$ is the value extracted from data using such $f_{\text{opt}}$, then $M_1$ can be used to reevaluate $f_{\text{opt}}$, and so on. The resulting iterative procedure is seen to be essentially equivalent to the optimization involved in the maximal likelihood method.

The beautiful aspect of (5) is that the quadratic nature of the second term on the r.h.s. of (5) (for which an explicit analytical expression is available [1]) means that $f_{opt}$ need not be known exactly: working with an approximation $f_{q\text{-}opt}$ to the optimal moment $f_{opt}$ may be practically sufficient. The approximation is then called *quasi-optimal observable*. Note that one may be able to construct satisfactory quasi-optimal observables even in situations when the method of maximal likelihood is not applicable, e.g. when the probability density is only known in the form of a MC generator rather than analytical formula.

---

[1] At this point it is worth emphasizing that parameter estimation is a more adequate interpretation of HEP data processing than event selection which — one tends to forget — is only an auxiliary tool.

As a typical example, consider measurements of a process mediated by an intermediate particle (cf. [2]; another example is the top quark search in the all-jets channel at D0 [7]). Then the probability distribution is represented as follows (remember that the overall normalization plays no role in this):

$$\pi(\mathbf{P}) = \pi_{bg}(\mathbf{P}) + g^2 \pi_{signal}(\mathbf{P}). \qquad (6)$$

The first unknown parameter is $g^2$, usually corresponding to the intermediate particle's coupling. Other parameters (e.g. its mass $M$) may be hidden within $\pi_{signal}$.

The optimal observable for measurement of $g^2$ (effectively, of the production rate for this channel) is

$$F_{opt}(\mathbf{P}) \left( \propto \frac{\partial \ln \pi(\mathbf{P})}{\partial g^2} \right) = \frac{g^2 \pi_{signal}(\mathbf{P})}{\pi_{bg}(\mathbf{P}) + g^2 \pi_{signal}(\mathbf{P})} \leq 1. \qquad (7)$$

It is straightforward to generate an approximation for this using a MC generator. Note that Eq. (7) approaches 1 whenever the signal exceeds the background appreciably, and approaches 0 in the opposite case. Should $F_{opt}$ happen to be such as to take only values 0 and 1 then it would exactly be the selection criterion that Dr. Anipko et al. [2] devised their algorithm to seek.

However, $F_{opt}$ is a continuous function in practical situations. This means that the optimal way (i.e. one ensuring minimum errors) to measure the production rate for this channel is to evaluate the sum

$$\sum_i F_{opt}(\mathbf{P}_i) \qquad (8)$$

rather than count events selected with any 0/1 approximation to $F_{opt}$. We are forced to conclude that *any* conventional data processing procedure based on 0/1 selection cuts — no matter what kind of intelligence it relies upon — involves a loss of physical information — i.e. yields systematically higher errors compared with the procedure based on the optimal observable.

How can one estimate the losses? Given that the hypersurface which separates the 0 and 1 regions in the space of events is mostly regular, and so has codim=1, it is sufficient to examine a one-dimensional situation. So let $\mathbf{P} \in [0,1]$ and $\pi(\mathbf{P}) = \mathbf{P}$. Compare the fluctuations of two observables, $f_{cont}(\mathbf{P}) = \mathbf{P}$ and $f_{cut}(\mathbf{P}) = \text{IF } \mathbf{P} < \frac{1}{2} \text{ THEN } 0 \text{ ELSE } 1 \text{ END}$. The second one corresponds to the usual cuts whereas the first one can be an optimal observable. For the statistical errors in the two cases, one finds $\sigma_{cut}/\sigma_{cont} = 1.6$ — a substantial factor sufficient to transform a $3\sigma$ signal into a $5\sigma$ discovery (or vice versa)! This is an optimistic case (or pessimistic, depending on the viewpoint). But it does indicate the range of possibilities.

Next, suppose one wishes to measure, say, the mass $M$ of the intermediate particle which is responsible for the production channel being studied. The corresponding optimal observable is represented as follows (the interesting factorized form is new compared with the earlier discussions):

$$f_{opt,M}(\mathbf{P}) = \frac{g^2 \partial_M \pi_{signal}}{\pi_{bg} + g^2 \pi_{signal}} = \boxed{F_{opt}(\mathbf{P}) \times \partial_M \ln \pi_{signal}(\mathbf{P})}. \qquad (9)$$

Again, the factorized form on the r.h.s. nicely corresponds to the conventional procedures: the first factor corresponds to event selection cuts for this channel and the second factor, to the specific measurement with the selected events.

Another example is as follows. One can construct optimal observables on events reduced to a few parameters using e.g. a jet algorithm. Then the quality of the resulting measurement procedure depends on the reduction (jet) algorithm used and one could employ the quantity $\langle f_{opt}^2 \rangle$ that emerges in (5) to compare different jet algorithms: the larger this number, the smaller the error of the measurement of $M$, and the better the jet algorithm. This number can also be used to control the tradeoffs between quality and speed involved in various optimizations. A comparison of this kind has been undertaken in [8]. Referring the reader for details to [8], I only cite the finding that the optimal jet definition introduced in [4] proves to be equivalent on this test to the $k_T$ algorithm [9], with both appreciably better than the JADE algorithm [10]. Furthermore, a simple optimization makes the implementation of the optimal jet definition reported in [11] more than twice as fast as any of the other two algorithms without noticeable loss in quality.

To conclude, the method of quasi-optimal observables combines:
- ✓ the simplicity of use of the method of moments;
- ✓ the optimal quality of results of maximal likelihood;
- ✓ algorithmic flexibility;
- ✓ a deterministic method to replace and improve upon event selection procedures based on cuts.
- ✓ It also meshes well with advanced theoretical calculations (allowing theorists to use arbitrarily singular expressions for higher order corrections to probability distributions — once quasi-optimal observables have been agreed upon with experimentalists);
- ✓ and offers a complete analytical control over one's optimizations etc. (via the quantity $\langle f_{opt}^2 \rangle$).

The next task is to engineer good software implementations of the method.

I thank E.Jankowski for several discussions and A.Czarnecki for the hospitality at the University of Alberta, Edmonton, where some of this work was done. This work was supported in parts by the NATO grant PST.CLG.977751 and the Natural Sciences and Engineering Research Council of Canada.

*Notes added*

After the first and second postings of this text, first Dr. A. Soni and then Drs. M. Diehl, O. Nachtmann and F. Nigel notified me of some relevant earlier publications. I am particularly indebted to Dr. Diehl for kindly sending me copies of the papers published in the early 90s that are unavailable on-line and in Russian libraries. The summary below rectifies some incorrect statements made in Notes added in the second posting of this text.

It turns out, the important special case of linear dependence on the measured parameter, Eqs. (6), (7), has a rather rich history of ten years — although it has been remaining a sort of esoteric knowledge in the community of specialists in weak and anomalous couplings.

Ref. [12] derived the corresponding optimal observable using orthogonality in Hilbert spaces familiar from quantum mechanics.

An interesting paper [13] dealt with the method of maximal likelihood in the case of the probability distribution (6) and pointed out that the likelihood function depends only on the real variable $\omega = \pi_{signal}(\mathbf{P})/\pi_{bg}(\mathbf{P})$, therefore, all the information on the parameter being estimated is carried by $\omega$. However, Ref. [13] stopped short of reformulating the estimation procedure in terms of the method of moments. As a kind of test of whether or not ref. [13] crossed the discovery line, I note the following: Whereas it would have been immediate to make use of the result of [12] in my work on jet observables [14] (and it would have saved me a lot of trouble then and later, even if the result is not general), it would be hardly possible to make a similar use of [13]; also, the problem was formulated in [12] in such a way that I can easily imagine myself replacing the derivation of [12] with the more general one [1] to arrive at the general result (4), but I cannot imagine ref. [13] as such a starting point.

An extension to several parameters, and with inclusion of quadratic terms (within the ideological framework of [12]) was presented and systematically studied in [15], including the connection with the FFRC inequality [16]. It resulted in a body of work on measurement of (anomalous) trilinear couplings (e.g. [17], [18]).

A generalization similar to that of [15] was achieved in [19].

Note that if the dependence on the parameter being estimated is essentially non-linear (as would be the case e.g. with masses, decay widths, etc.), the general formulae (4) and (9) have to be invoked.

It is rather amusing that the citations of [12], [15], [19] are strongly correlated with the keywords "CP violation", "trilinear gauge couplings", "anomalous three gauge boson couplings". The corresponding community seems to have no appreciable intersection with that specializing in jet algorithms or in the construction of cuts using complicated algorithmic machinery (as represented e.g. at the ACAT/AIHENP workshops).

Still, I find it strange that little, if any, effort seems to have been expended to decouple the results of refs. [12], [15], [19] from the rather special physical problematics and to make them known beyond the community of weak-couplings specialists, as those formulae deserved.

## References

[1] F.V.Tkachov, physics/0001019, Part. Nucl., Letters, **2[111]** (2002) 28.
[2] D. Anipko, *Search of natural cuts in seeking of new physics phenomena*, talk at ACAT'02, ID=139.
[3] F.V.Tkachov, physics/0108030, in: *XI Int. School "Particles and Cosmology", Baksan Valley, 18-24 April 2000*, INR RAS, Moscow.
[4] F.V. Tkachov, hep-ph/9901444, Int. J. Mod. Phys. **A17** (2002) 2783.
[5] T. Kuhn, *The Structure of Scientific Revolutions,* Chicago Univ. Press, 1962.
[6] F.V. Tkachov, hep-ph/0202033 [talk at CPP'2001, Tokyo, 28-30 Nov 2001]
[7] P. Bhat, H. Prosper and S. Snyder, hep-ex/9809011, IJMP **A13** (1998) 5113.
[8] E. Jankowski, D.Yu. Grigoriev and F.V.Tkachov, in preparation.
[9] S. Catani et al., Phys. Lett. **269B** (1991) 432.
[10] JADE Collaboration (W. Bartel et al.), Z. f. Physik **C33** (1986) 23.
[11] D.Yu. Grigoriev and F.V. Tkachov, hep-ph/9912415 [in QFTHEP'99, Moscow, May 27 - June 2 1999, MSU-Press, Moscow].
[12] D. Atwood and A. Soni, Phys. Rev. **D45** (1992) 2405.
[13] M. Davier, L. Duflot, F. Le Diberder and A. Rougé, Phys. Lett. **B306** (1993) 411.
[14] F.V. Tkachov, Phys. Rev. Lett. **73** (1994) 2405.
[15] M. Diehl and O. Nachtmann, Z. Phys. **C62** (1994) 397;
M. Diehl and O. Nachtmann, Eur. Phys. J. **C1** (1998) 177, hep-ph/9702208;
M. Diehl, O. Nachtmann and F. Nagel, hep-ph/0209229.
[16] R.A. Fisher, Proc. Camb. Phil. Soc. **22** (1925) 700-725;
M. Frechet, Rev. Intern. de Stat. 1943, 182;
C.R. Rao, Bull. Calcutta Math. Soc. **37** (1945) 81-91;
H. Cramer, Aktuariestidskrift **29** (1946) 458-463.
[17] G.J. Gounaris and C.G. Papadopoulos, Eur. Phys. J. **C2** (1998) 365, hep-ph/9612378.
[18] G.K. Fanourakis, D. Fassouliotis, S.E. Tzamarias, Nucl. Instrum. Meth. **A412** (1998) 465, hep-ex/9711014;
G.K. Fanourakis et al., Nucl. Instrum. Meth. **A430** (1999) 465, hep-ex/9812002.
[19] J.F. Gunion, B. Grzadkowski and X.-G. He, Phys. Rev. Lett. **77** (1996) 5172.